\documentclass[%
 reprint,
 superscriptaddress,
 amsmath,amssymb,
 prl,
floatfix,
]{revtex4-1}

\usepackage{graphicx}
\usepackage{dcolumn}
\usepackage{bm}

\usepackage{color}
\usepackage{braket}

\begin{document}

\title{Electron-phonon interaction and longitudinal-transverse phonon splitting in doped semiconductors}

\author{Francesco Macheda}
\affiliation{Istituto Italiano di Tecnologia, Graphene Labs, Via Morego 30, I-16163 Genova, Italy}%
\author{Paolo Barone}%
\affiliation{CNR-SPIN, Area della Ricerca di Tor Vergata, Via del Fosso del Cavaliere 100,
I-00133 Rome, Italy}%
\affiliation{Dipartimento di Fisica, Università di Roma La Sapienza, Roma, Italy }%
\author{Francesco Mauri}
\affiliation{Dipartimento di Fisica, Università di Roma La Sapienza, Roma, Italy }%
\affiliation{Istituto Italiano di Tecnologia, Graphene Labs, Via Morego 30, I-16163 Genova, Italy}%

\newcommand{\citeSupp}[0]{Note1}

\begin{abstract}
We study the effect of doping on the electron-phonon interaction and on the phonon frequencies in doped semiconductors, taking into account the screening in presence of free carriers at finite temperature. We study the impact of screening on the Fr\"ohlich-like vertex and on the long-range components of the dynamical matrix, going beyond the state-of-the-art description for undoped crystals, thanks to the development of a computational method based on maximally localized Wannier functions. We apply our approach to cubic silicon carbide, where in presence of doping the Fr\"ohlich coupling and the longitudinal-transverse phonon splitting are strongly reduced, thereby influencing observable properties such as the electronic lifetime.
\end{abstract}

\maketitle

\footnotetext[1]{ see Supplementay Information at.}

The calculation of the electron-phonon interaction (EPI)~\cite{ziman2001electrons,ziman2001electrons,Grimvall1981TheEI,schrieffer1999theory,mahan1990many,RevModPhys.89.015003} is of crucial importance in order to predict and interpret, e.g., transport experiments~\cite{PhysRevB.98.201201,PhysRevB.102.094308,Macheda2020,PhysRevB.100.085204,PhysRevResearch.3.043022,Lee2020,Ponce2020,PhysRevB.94.201201,macheda2021ab,PhysRevB.97.045201,PhysRevResearch.2.033055}, excited carriers relaxation~\cite{PhysRevB.84.075449,Harb2006,Betz2013} and superconductivity~\cite{PhysRev.108.1175,schrieffer1999theory}. The precise evaluation of the EPI has become possible thanks to density functional theory (DFT)~\cite{PhysRev.136.B864,PhysRev.140.A1133} and density functional perturbation theory (DFPT)~\cite{RevModPhys.73.515}, 
alongside with Wannier interpolation technique~\cite{RevModPhys.84.1419,PhysRevB.76.165108,PhysRevB.82.165111} and the progress of \textit{ab-initio} computational infrastructures~\cite{doi:10.1063/5.0005082,Pizzi2020,PONCE2016116}.
Despite recent advances, the long-range Coulomb-mediated contributions to both phonons and EPI in the long-wavelength limit are usually accounted for via semi-empirical approaches describing the macroscopic electric fields created by oscillating strain fields or longitudinal optical phonons~\cite{RevModPhys.73.515,yu2010fundamentals}. As such, they should be limited to semiconductors and insulators, as they neglect screening effects due to the presence of free carriers. A more precise treatment has been proposed for the long-range part of the polar-optical EPI (Fr\"ohlich-like term)~\cite{PhysRevB.92.054307,PhysRevLett.115.176401,PhysRevLett.125.136601,PhysRevB.102.125203,PhysRevB.92.054307}, based on the microscopic theory firstly developed by Vogl but still limited to the undoped setup~\cite{PhysRevB.13.694}. In doped semiconductors, the presence of even tiny fraction of carriers may significantly alter the electronic dielectric properties in the relevant long-wavelength limit, resulting in effectively screened macroscopic electric fields and hence in a reduction of piezoelectric/Fr\"ohlich-like EPI as well as of the longitudinal optical-transverse optical (LO-TO) mode splitting. 
The lack of an explicit formulation of the dielectric response beyond the undoped setup hindered so far a precise assessment of the limits of validity of such approximation, usually adopted when evaluating the long-range contributions to EPI and LO-TO splitting from first principles. This may lead to their severe overestimation in doped semiconductors and to incorrect evaluation of related physical properties as, e.g., electronic lifetimes and scattering rates.
The potential impact of these effects has been recently discussed, within a semi-empirical approach based on Thomas-Fermi theory of screening, in doped half-Heusler semiconductors~\cite{Ren2020}, whose non-monotonous carrier mobility has been related to doping-induced changes of polar optical phonon scattering and to the collapse of LO-TO splitting.

In this work we overcome the above described limitations and quantify the implications of the presence of free carriers in a doped semiconductor at finite temperature. Within a linear-response and dielectric matrix formulation, we derive the general expressions of the long-range contributions to the dynamical matrix and EPI, that reduce to well-established formulae in the semiconducting regime and thus allow for controlled approximations of screening effects in appropriate doping-temperature regimes. Operatively, we propose a precise, stable and fast technique based on first principles calculations and Wannier interpolation, building on our general formulation and supported by computationally exact evaluation of both the screened and unscreened charge response, in order to obtain the correctly screened phonon frequencies and EPI on fine grids in the BZ.

We extend the approach developed in the seminal works of Pick, Cohen and Martin~\cite{PhysRevB.1.910}, Vogl~\cite{PhysRevB.13.694} and Stengel~\cite{PhysRevB.88.174106} to the case of doped semiconductors and metals. In this general treatment, it is still convenient to split the dynamical matrix and the EPI as the sum of a short-range component (SRC or S) and a long-range one (LRC or L), as routinely done for the particular case of semiconductors. Within a static linear-response approach, the LRCs can be recasted as~\cite{SI}:

\begin{align}
& C^{\textrm{L}}_{ss',\alpha \beta}(\mathbf{q})= \frac{4\pi e^2}{V}Z^{\textrm{c.c.}}_{s,\alpha}(\mathbf{q}) \bar Z_{s',\beta}(\mathbf{q}) \label{eq:1},\\
& g^{\textrm{L}}_{\nu, mm'}(\mathbf{k},\mathbf{q})= i\frac{4\pi e^2}{Vq} \sum_{s\alpha}Z_{s,\alpha}(\mathbf{q})  e^{\nu}_{s,\alpha}(\mathbf{q})l^{\nu}_{\mathbf{q}}\left( \frac{M_0}{M_s}\right)^{1/2} \times \nonumber \\
& \times \braket{u_{m\mathbf{k+q}}|u_{m'\mathbf{k}}}, \label{eq:2}
\end{align}
where $\textrm{c.c.}$ stands for complex conjugate, $V$ is the volume of the unit cell, $s,s'$ are atomic indexes, $\alpha,\beta$ are Cartesian indexes, $\nu$ is the phonon branch index, $m',m$ are electronic band indexes, $\mathbf{k}$ and $\mathbf{q}$ are the electronic and phonon wavevectors, $\mathbf{e}^{\nu}(\mathbf{q})$ is the phonon eigenvector, $l^{\nu}_{\mathbf{q}}=[\hbar/(2M_0\omega_{\nu\mathbf{q}})]^{1/2}$ is the phonon displacement amplitude, $M_0$ is an arbitrary reference mass, $M_s$ is the mass of the atom $s$ and $u$ are the periodic parts of the Bloch functions. The central quantities are $Z_{s,\alpha}(\mathbf{q})$ and $\bar Z_{s,\alpha}(\mathbf{q})$, the macroscopically \textit{screened} and \textit{unscreened} effective charges, respectively, defined as
\begin{figure}
\includegraphics[width=1\columnwidth]{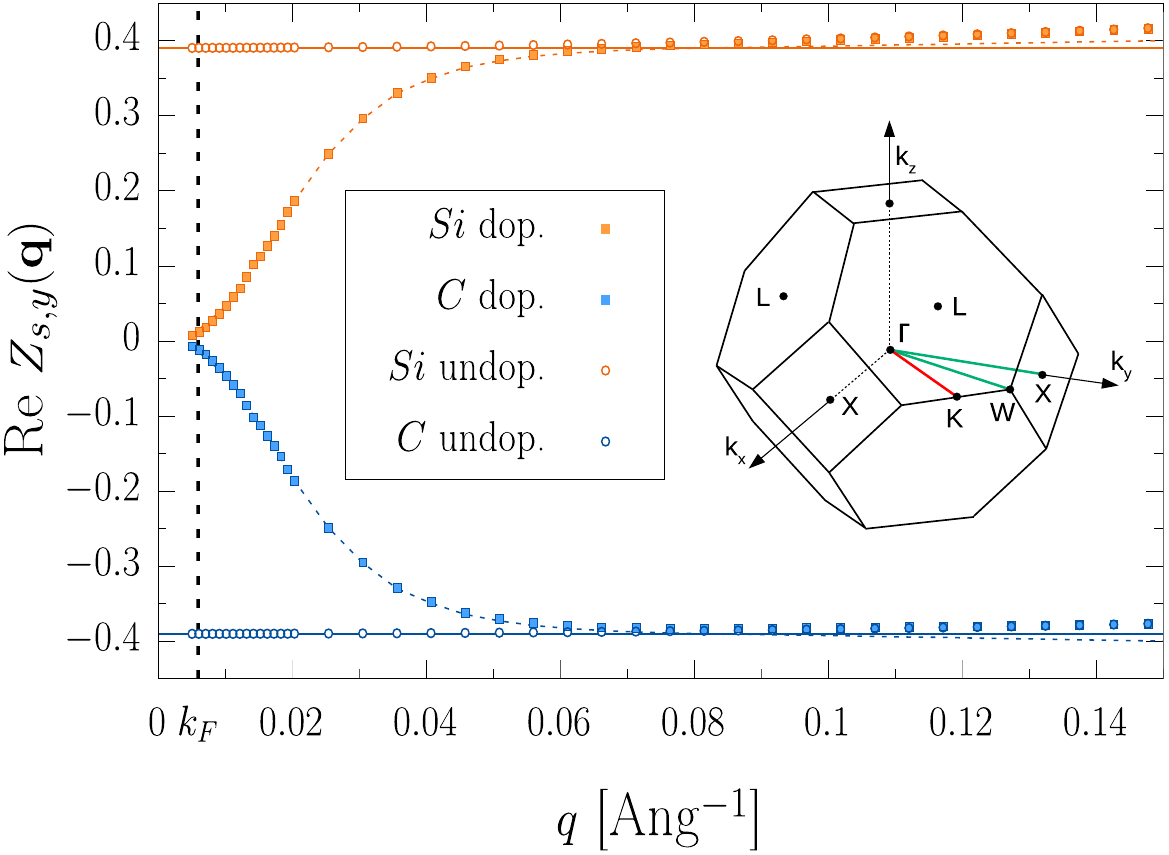}
\caption{Real part of $Z_{s,y}(\mathbf{q})$ as a function of $q$ along the line $\mathbf{q}=\left(q/\sqrt{2},q/\sqrt{2},0\right)$ (chosen for symmetry considerations on the effective charge tensors for the cubic case~\cite{SI} and displayed in the BZ as a red line) for both the atoms in 3C-SiC in the undoped case (empty circles) and in the doped case with an hole carrier concentration of $p=4.827\times 10^{15}$cm$^{-3}$ (full squares). The continuous lines are drawn using Eq. \ref{eq:4}, while the dashed ones are drawn using Eq. \ref{eq:5}. The black vertical line is drawn at the effective Fermi momentum $k_F$, as explained in the text.}
\label{fig:1}
\end{figure}

\begin{align}
Z_{s,\alpha}(\mathbf{q})=\epsilon^{-1}(\mathbf{q})\bar Z_{s,\alpha}(\mathbf{q})=+i\frac{V}{eq}\delta \rho^{\textrm{ind}}_{s,\alpha}(\mathbf{q})+\frac{q_{\alpha}}{q}Z_{s}, \label{eq:scrcharge}
\end{align}
where $Z_{s,\alpha}$ has been introduced in Ref.~\cite{Senga2019} to model EELS spectra, $Z_s$ is the ionic charge, $\epsilon^{-1}(\mathbf{q})$ the macroscopic scalar inverse dielectric function and $\delta \rho^{\textrm{ind}}_{s,\alpha}(\mathbf{q})$ the induced charge density produced by a collective displacement of wavevector $\mathbf{q}$ of the $s$ atoms along $\alpha$, i.e., $\upsilon_{s,\alpha}(\mathbf{q})=\lambda_{\alpha} \sum_p e^{i\mathbf{q}\cdot(\mathbf{R}_p+\boldsymbol{\tau}_s)}$ where $\boldsymbol{\lambda}$ is the magnitude of the atomic displacement; the absence of $e^{i\mathbf{q}\cdot\boldsymbol{\tau}_s}$ with respect to Eq. 2 of Ref.~\cite{Senga2019} comes from a different definition of the phonon eigenvectors~\cite{[{}][{Within our conventions the displacement of the atom at $\mathbf{R}_p+\boldsymbol{\tau}_s$, induced by a phonon of wavevector $\mathbf{q}$ and branch index $\nu$, is written as $\mathbf{e}^{\nu}_{s}(\mathbf{q})e^{i\mathbf{q}\cdot(\mathbf{R}_p+\boldsymbol{\tau}_s)}$}]phdispl}. Both $Z_{s,\alpha}(\mathbf{q})$ and $\bar Z_{s,\alpha}(\mathbf{q})$ can be obtained via DFPT{, the latter by imposing that the macroscopic component of the electrostatic potential vanishes~\cite{PhysRevB.88.174106,SI}}.
It can be shown that $\bar Z_{s,\alpha}(\mathbf{q})$ enjoys an expansion of the form (now expliciting the dependence on the carrier concentration $n$ at finite temperature $T$)

\begin{align}
\bar Z_{s,\alpha}(\mathbf{q},n,T)=\frac{i}{q}M_{s,\alpha}(n,T)+\frac{q_{\beta}}{q}Z^*_{s,\alpha\beta}(n,T)+ \nonumber\\
-\frac{i}{2} \frac{q_{\beta}}{q}q_{\gamma}Q_{s,\alpha\beta\gamma}(n,T)+\ldots \,. \label{eq:3}
\end{align}
\begin{figure}
\includegraphics[width=\columnwidth]{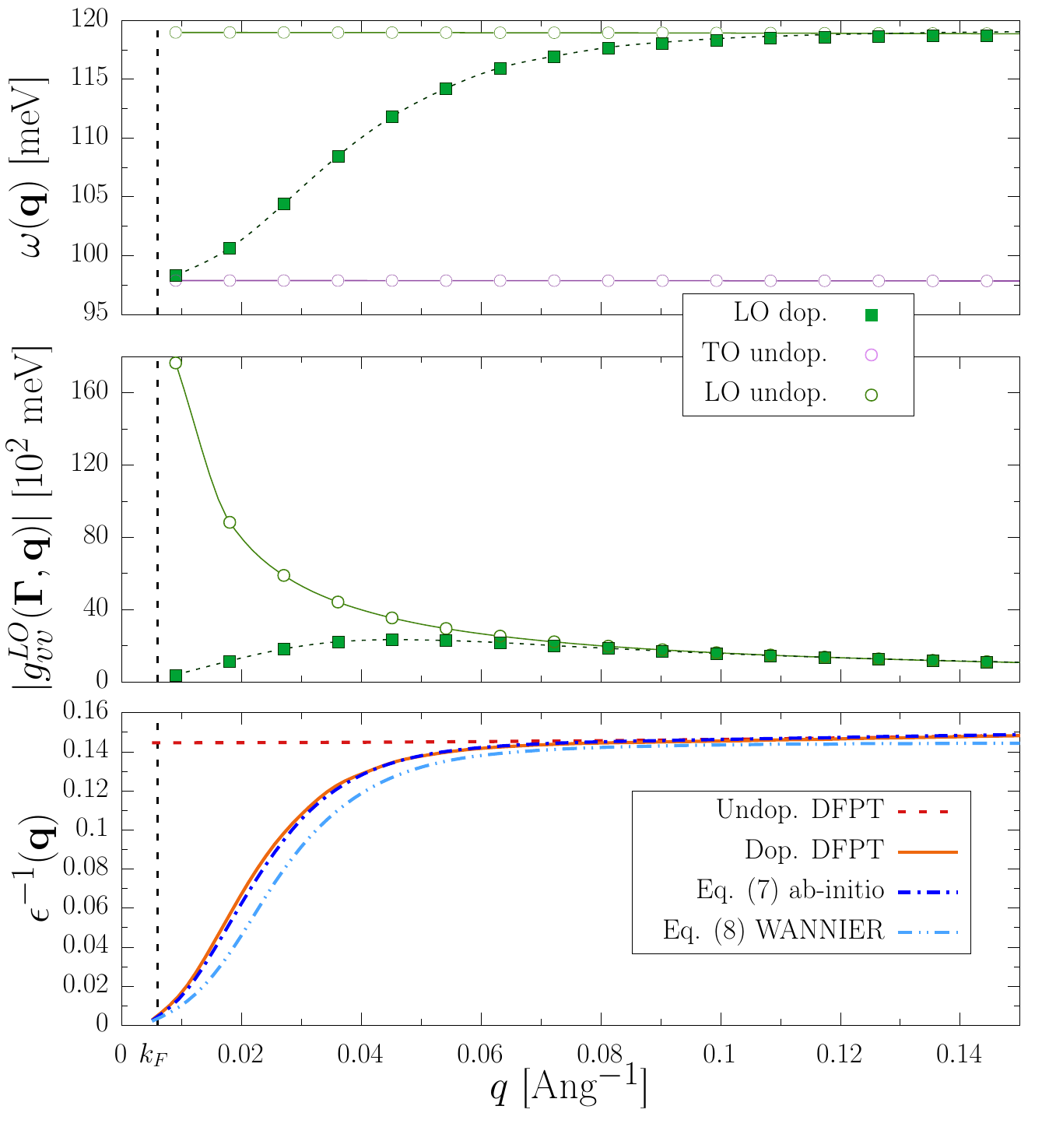}
\caption{Upper) Phonon frequencies from first-principles in the undoped and doped setup (empty circles and full squares), and their interpolation (continuous and dashed lines). The TO mode is presented only for the undoped case. Central) Same comparison for the EPI. The electronic indexes of $g$ are both referred to the highest valence band. Lower) Comparison of the macroscopic inverse dielectric screening computed from first principles in the undoped (red line) and doped setups (orange line) with the one obtained via Eq. \ref{eq:6} using \textit{ab-initio} energies and overlaps (blue line) and with Eq. \ref{eq:6.1} using interpolated energies (light-blue line).}
\label{fig:2}
\end{figure}
The first constant tensor of the expansion is related to the presence of a Fermi energy shift in systems that allows it by symmetry reasons (see Eq. (79) of Ref.~\cite{RevModPhys.73.515} and~\cite {SI}), the second with Born effective charges, the third with dynamical effective quadrupoles and so on (sums over repeated indexes are intended).

For an undoped semiconductor, the term $M_{s,\alpha}$ is zero~\cite{SI}; also, Eqs. (\ref{eq:1})-(\ref{eq:2}) are non-analyitical because $\lim_{\mathbf{q}\rightarrow 0} \epsilon^{-1}(\mathbf{q})=q^2/\left( \mathbf{q} \cdot \overleftrightarrow \epsilon^{\infty} \cdot \mathbf{q} \right)$, where $\overleftrightarrow \epsilon^{\infty}$ is the high-frequency static macroscopic dielectric tensor~\cite{PhysRevB.43.7231}.
Combining this limiting expression with the lowest orders in the expansion Eq. (\ref{eq:3}) (typically only Born effective charges, see Refs.~\cite{PhysRevResearch.3.043022,PhysRevLett.125.136601,PhysRevLett.125.136602} for recent developments including dynamical effective quadrupoles), one recovers the usual expressions for LRCs~\cite{SI}. The interpolation of dynamical matrix and EPI near the BZ center is then performed by first interpolating, for a general $(\mathbf{\bar k},\mathbf{\bar q})$, the differences $C^{\textrm{S}}=C-C^{\textrm{L}}$ and $g^{\textrm{S}}=g-g^{\textrm{L}}$ via Fourier or Wannier interpolation~\cite{PhysRevB.76.165108}; as LRCs are explicitly known only for $\mathbf{q}\rightarrow 0$, their expression is extended on the whole BZ applying Ewald summation techniques with the correct limiting values~\cite{PhysRevB.50.13035,PhysRevB.92.054307}. Finally, the LRCs of Eqs. (\ref{eq:1})-(\ref{eq:2}) are evaluated at $(\mathbf{\bar k},\mathbf{\bar q})$ and are added to the SRCs to obtain the correctly interpolated quantities. 

For doped semiconductors and metals, it would seem that the above interpolating procedure is not required as the dependence on $n$ and $T$ makes Eqs. (\ref{eq:1})-(\ref{eq:2}) analytical around $\Gamma$ due to the regularization of the scalar inverse dielectric matrix for small $\mathbf{q}$.
Nonetheless, in the small-doping limit relevant, e.g., for intrinsic transport coefficients, the regularization region in reciprocal space is vanishingly small and does not guarantee that the dynamical matrix and the EPI can be interpolated without precautions. Consequently, the interpolation of the phonon frequencies and the EPI on a generic $(\mathbf{\bar k},\mathbf{\bar q})$ point still requires techniques that rely on the subtraction and addition of the asymptotic formula for the LRCs, which in principle have to be calculated for each $n$ and $T$. In practice, one can compute the EPI and dynamical matrices from \textit{ab-initio} and extract the SRCs on a coarse grid of points for the undoped case at 0K, interpolate the SRCs in $(\mathbf{\bar k},\mathbf{\bar q})$, and then add the LRCs of Eqs. (\ref{eq:1})-(\ref{eq:2}) on the same point but now including the correct $n$ and $T$ dependences. A main simplification comes from the fact that, as we will show later for a representative example, for sufficiently small $n$ and $T$ and for polar systems where $M_{s,\alpha}=0$ for symmetry reasons, at the leading order in $\mathbf{q}$ we can approximate $\bar Z_{s,\alpha}(\mathbf{q},n,T) \approx \bar Z_{s,\alpha}(\mathbf{q},n=0,T=0)$ .

Given the above, to evaluate the leading order LRCs in doped semiconductors we can use Eqs. (\ref{eq:1})-(\ref{eq:2}) with the effective charge tensors of Eq. (\ref{eq:3}) deduced 
in the undoped $T=0$ setup and $\epsilon^{-1}(\mathbf{q},n,T)$; the computation of $\epsilon^{-1}(\mathbf{q},n,T)$ can be performed exactly in the DFPT framework (see Eq. (10) and the related discussion in Ref.~\cite{PhysRevB.91.165428}) with a relatively small computational cost \cite{SI}. 
Using crystalline symmetries, the leading order of $\epsilon^{-1}(\mathbf{q},n,T)$ can be evaluated from calculations performed on a number of lines which is equal to the number of independent components of the $\overleftrightarrow \epsilon^{\infty}$ tensor~\cite{SI}.

In order to validate the above described theoretical approach, we numerically study 3C-SiC---one of the promising wide-gap semiconductors for power electronics~\cite{CHOW2006112}---as a prototypical example of polar material where phonon frequencies and EPI are heavily influenced by doping; in this system the term $M_{s,\alpha}$ is zero for symmetry reasons. To perform the first-principles calculations we use a private version of QUANTUM ESPRESSO~\cite{doi:10.1063/5.0005082}, as developed in Ref.~\cite{Senga2019}. We use PBE-GGA functionals~\cite{PhysRevLett.77.3865} and norm-conserving pseudopotentials, and we sample the BZ with telescopic grids that accurately describe the region around the chemical potential ~\cite{SI}. We use an hole doping concentration of $p=4.827\times 10^{15}$cm$^{-3}$ (roughly the lower limit of concentrations for p-type 3C-SiC transport experiments~\cite{doi:10.1063/1.338211}) by reducing the total number of electrons in the system while introducing a negative uniform jellium background to preserve charge neutrality. The Wannier interpolation is performed using a private version of EPW~\cite{PONCE2016116} (more information at~\cite{SI}).

We start by studying $Z_{s,\alpha}(\mathbf{q})$ for the undoped and the doped setups at zero temperature, simulated in the latter case with a Gaussian smearing of $10^{-5}$ Ry. We present in Fig. \ref{fig:1} the fully \textit{ab-initio} calculation of $ \textrm{Re}Z_{s,y}(\mathbf{q})$ along a representative line in reciprocal space, comparing it with the leading order expansions~\cite{SI}
\begin{align}
Z_{s,y}(\mathbf{q})=Z^*_{s}/\sqrt{2}/\epsilon^{\infty}, \label{eq:4}\\
Z_{s,y}(\mathbf{q},n,T)= \epsilon^{-1}(\mathbf{q},n,T) Z^*_{s}/\sqrt{2}.
\label{eq:5}
\end{align}
Here $Z^*_{s,\alpha\beta}=Z^*_{s}\delta_{\alpha\beta}$ and $\overleftrightarrow \epsilon^{\infty}_{\alpha\beta}=\epsilon^{\infty} \delta_{\alpha\beta}$ are computed with DFPT at $\mathbf{q}=\mathbf{0}$ in the undoped setup, while $\epsilon^{-1}(\mathbf{q},n,T)$ is computed using DFPT for each $\mathbf{q}$ point in the doped case. We immediately notice that on the scale of the effective Fermi momentum $k_F$, defined as the wavevector where the hole occupation halves from its value at the top valence band, the screening of the effective charges is practically complete. Secondly, we notice that the agreement between calculations and Eqs. (\ref{eq:4}) and (\ref{eq:5}) is excellent at small $\mathbf{q}$, with small differences arising at larger wavevectors that are ascribable to higher order terms in $\bar{Z}_{s,y}(\mathbf{q})$. Notably, in Eq. (\ref{eq:5}) we have used the Born effective charge tensor computed in the undoped setup, signalling that the effect of doping enters predominantly in the scalar macroscopic inverse dielectric function, as anticipated.

We are now in position to perform Fourier and Wannier interpolation of phonon frequencies and EPI using Eqs. (\ref{eq:1})-(\ref{eq:2}), and evaluate its quality for the undoped and the doped cases in the upper and central panels of Fig. \ref{fig:2}. The agreement between \textit{ab-initio} and interpolated quantities, not only for the undoped case and in the region where the LRCs dominate, but on the whole BZ region, is striking. From a physical point of view, it is evident that doping effects are very relevant for the asymptotic behaviour of both the phonon frequencies and the EPI in the $\mathbf{q}\rightarrow \mathbf{0}$ limit: both the LO-TO splitting and the Fr\"{o}hlich coupling are substantially suppressed on the scale of $k_F$, i.e. in the region important for e.g. transport properties.

A diriment question regarding the methodology presented here is whether we can obtain the macroscopic dielectric function $\epsilon^{-1}(\mathbf{q},n,T)$ without resorting to first-principles calculations but sill preserving the qualitative trends presented in Fig. \ref{fig:2}, thus providing a fast method to evaluate the dielectric screening for a whole range of carrier concentrations and temperatures. To reach our goal, we use the RPA relation $\epsilon(\mathbf{q},n,T)=1-4\pi e^2/q^2 \chi^0(\mathbf{q},n,T)$, and $\epsilon^{-1}(\mathbf{q},n,T)\approx 1/\epsilon(\mathbf{q},n,T)$; the latter relation is a reasonable approximation that neglects local fields~\cite{SI}, typically giving corrections of order $10\%$ on $\epsilon^{\infty}$~\cite{PhysRevB.23.6615,PhysRevB.33.7017,refId0}. We evaluate $\chi^0(\mathbf{q},n,T)$ 
as done in Ref.~\cite{PhysRevMaterials.5.024004}, splitting it into interband and intraband contributions, eventually leading to
\begin{align}
& \epsilon^{-1}(\mathbf{q},n,T)\approx \frac{1}{\frac{1}{\epsilon^{-1,\textrm{undop}}(\mathbf{q})}-4\pi e^2/q^2 \delta \chi^0(\mathbf{q},n,T)}, \label{eq:6} \\
& \delta \chi^0(\mathbf{q},n,T) = \frac{2}{V}\sum_{mm'\mathbf{k}}  \frac{\delta f_{m\mathbf{k}}-\delta f_{m'\mathbf{k+q}}}{\epsilon_{m\mathbf{k}}-\epsilon_{m'\mathbf{k+q}}} |\braket{u_{m\mathbf{k}}|u_{m'\mathbf{k+q}}}|^2, 
\nonumber
\end{align}
\begin{figure}
\includegraphics[width=1\columnwidth]{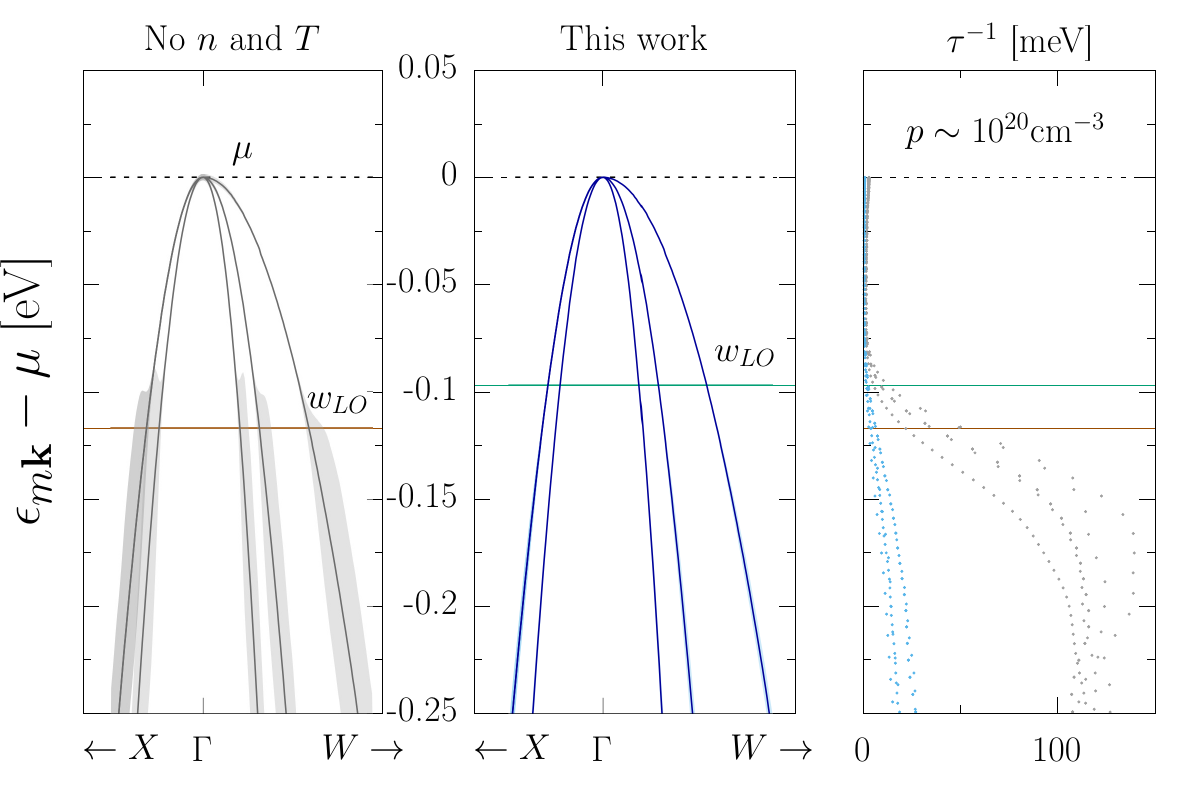}
\caption{Comparison of inverse lifetime scattering computed along high symmetry lines using Eq. (\ref{eq:7}) where the LRCs are computed using Eqs. (\ref{eq:1})-(\ref{eq:2}) with and without $n$ and $T$ dependence for a hole concentration of $p\sim10^{20}$cm$^{-3}$: the inverse lifetimes due to LO phonons are given in absolute magnitude in the right panel, and as linewidths in the left and central ones. The light blue dots and the central panel are computed with the method proposed in this work, while grey dots and the left panel are obtained using state-of-the-art methods. The $\mathbf{k}$ point path of the calculation is shown in the inset of Fig. \ref{eq:1} with green lines. The brown and green horizontal lines are placed at $-\omega_{LO}(\mathbf{q}=0)$, while the black dotted line indicates the position of the chemical potential.}
\label{fig:4}
\end{figure}
where $\delta f=f^{\textrm{dop}}-f^{\textrm{undop}}$, $f$ being the electronic statistical distribution, and 
$\epsilon^{-1,\textrm{undop}}(\mathbf{q},T)\approx \epsilon^{-1,\textrm{undop}}(\mathbf{q})$ can be computed just once \textit{ab-initio}; we have also supposed that DFT energies and overlaps can be approximated with the ones computed in the undoped setup. A further practical simplification comes from neglecting the $\mathbf{q}$ dependence of the dielectric response using $1/\epsilon^{-1,\textrm{undop}}(\mathbf{q})\approx \epsilon^{\infty}$, neglecting the sum over conduction states, and taking $\braket{u_{m\mathbf{k}}|u_{m'\mathbf{k+q}}}=\delta_{mn}$~\cite{SI}:
\begin{align}
& \epsilon^{-1}(\mathbf{q},n,T)\approx \frac{1}{\epsilon^{\infty}-4\pi e^2/q^2 \delta \chi^0(\mathbf{q},n,T)}, \label{eq:6.1} \\
& \delta \chi^0(\mathbf{q},n,T) = \frac{2}{V}\sum_{m\mathbf{k}}  \frac{\delta f_{m\mathbf{k}}-\delta f_{m\mathbf{k+q}}}{\epsilon_{m\mathbf{k}}-\epsilon_{m\mathbf{k+q}}}, 
\nonumber
\end{align}
with
the sum over $m$ 
restricted to valence band only.
The result of this procedure is compared in the lower panel of Fig. \ref{fig:2} with the first-principles $\epsilon^{-1}(\mathbf{q},n,T)$. Despite the number of approximations involved, we notice that the agreement is qualitatively good, and it comes along with a great speed-up in computational cost since Eq. (\ref{eq:6.1}) can be evaluated very quickly on a large set of points. 

Finally, to quantify the impact that the correct descriptions of the LRCs of the EPI and phonon frequencies have on physical observables, we evaluate the electronic inverse lifetimes due to the interaction with LO phonons (which is most often found to be the dominant mechanism that limits the intrinsic carrier mobility at room temperature in polar semiconductors~\cite{PhysRevResearch.3.043022}), as evaluated in the Self Energy Relaxation Time Approximation (SERTA)~\cite{Ponce2020}
\begin{align}
\tau^{-1}_{m\mathbf{k}}&=\frac{2\pi}{\hbar} \sum_{m'} \int\frac{d\mathbf{q}}{V_{BZ}} |g_{\textrm{LO},m'm}(\mathbf{k},\mathbf{q})|^2 \times \nonumber \\
&\big[(1-f_{m'\mathbf{k+q}}+n_{\nu\mathbf{q}})\delta(\epsilon_{m\mathbf{k}}-\epsilon_{m'\mathbf{k+q}}-\hbar \omega_{\nu\mathbf{q}}) + \nonumber \\
&(f_{m'\mathbf{k+q}}+n_{\nu\mathbf{q}})\delta(\epsilon_{m\mathbf{k}}-\epsilon_{m'\mathbf{k+q}}+\hbar \omega_{\nu\mathbf{q}}) \big], 
\label{eq:7}
\end{align}
where $V_{BZ}$ is the volume of the BZ, and $n_{\nu\mathbf{q}}$ is the Bose-Einstein occupation factor of the phonon of frequency $\omega_{\nu\mathbf{q}}$. We compare the cases where the LRCs are interpolated i) using 
Eq. (\ref{eq:4}) (as currently done in most state-of-the-art first-principles calculations) and ii) using 
Eq. (\ref{eq:5}), thus including dielectric screening in a controlled approximation at the RPA level. Since the doping-temperature regimes 
with significant changes of screening properties depend critically on the material electronic properties, as a representative comparison we choose a setup tailored to highlight the differences between the two approaches. For sake of clarity, we fix the chemical potential $\mu$ at the valence-band top and set $T=300~$K, so that the numerical difference in the approaches can be ascribed only to differences in the screening; the corresponding carrier concentration is approximately $10^{20}$cm$^{-3}$.  As shown in Fig. \ref{fig:4},  discarding the doping dependence of the screening implies an overestimation of the LO contribution to the inverse scattering times up to a factor $\sim 5$ as soon as the onset of optical phonon scattering is reached ($\omega_{LO}$ in figure); moreover, in the doped case the LO contribution to the carriers lifetime at the top of the valence band is completely suppressed.

In conclusion, we have analysed the dependence on doping and temperature of EPI and phonon frequencies in doped semiconductors, within a linear-response dielectric-matrix formulation that allows for controlled approximations of the effect of electronic screening. We further propose a fast and accurate interpolation method, validated by fully \textit{ab-initio} calculations, that enables a quantitative analysis at a feasible computational cost. We have shown that neglecting free-carriers screening, as typically done in state-of-the-art computational approaches, may lead to inaccurate conclusions on transport properties arising from a substantial overestimation, in specific doping-temperature regimes, of scattering rates due to the Fr\"ohlich interaction. However, our general formulation is not limited to the latter, but applies to any EPI that are accessible within DFPT, i.e., to the lowest order in perturbation theory. The proposed approach for dealing with electronic screening lay the foundation for further extensions tackling other less conventional types of EPI, such as the vector coupling or the electron two-phonon scattering, that may play an important role in polar metals and doped quantum paraelectrics~\cite{GASTIASORO2020168107,PhysRevLett.126.076601,PhysRevB.105.125142,PhysRevB.94.224515}. Lastly, the extension to finite-frequencies dependence within a time-dependent DFPT approach can also give access to nonadiabatic effects on effective charges and, hence, on lattice dielectric properties~\cite{PhysRevB.103.134304, PhysRevLett.128.095901}, as well as to the vibrational contribution to the full macroscopic dielectric function that shapes the dynamical structure factor probed by EELS~\cite{Senga2019} and other inelastic scattering experiments~\cite{Sinha2001}.

\begin{acknowledgments}
We acknowledge the European Union’s Horizon 2020 research and innovation program under grant agreements no. 881603-Graphene Core3 and the MORE-TEM ERC-SYN project, grant agreement No 951215. We acknowledge that the results of this research have been achieved using the DECI resource \textit{Mahti CSC} based in Finland at https://research.csc.fi/-/mahti with support from the PRACE aisbl; we also acknowledge PRACE for awarding us access to Joliot-Curie Rome at TGCC, France. We thank Dr. Nicola Bonini and Dr. Thibault Sohier for useful discussions. 
\end{acknowledgments}

\bibliography{biblio}

\end{document}